\documentclass[11pt]{article}

\usepackage{jcapmod}
\usepackage{booktabs}
\usepackage[english]{babel}
\usepackage{amsmath,amssymb,amsbsy,amstext, amsthm}
\usepackage{bm}
\usepackage{graphicx}
\usepackage{amsfonts}
\usepackage{amssymb}
 \usepackage{exscale,relsize}
 \usepackage[makeroom]{cancel}
\usepackage{soul}
\usepackage{bm}
\RequirePackage{color}

\usepackage{colortbl}
\definecolor{rp}{cmyk}{0.2, 1, 0.6, 0}
\definecolor{green2}{cmyk}{0, 1, 0.5, 0}
\definecolor{lightgreen}{cmyk}{0.2, 0, 0.2, 0.2}
\definecolor{lightgray}{cmyk}{0.1,0.2,0,0.1}
\definecolor{lightgray2}{cmyk}{0.4,0.4,0,0.8}
\definecolor{black}{cmyk}{1.0,1.0,1.0,1.0}

\allowdisplaybreaks[1]

\usepackage{colortbl}
\definecolor{lightgreen}{cmyk}{0.2, 0, 0.2, 0.2}
\definecolor{lightgray}{cmyk}{0.1,0.2,0,0.1}
\definecolor{lightgray2}{cmyk}{0.1,0.1,0,0.1}

\makeatletter
\newlength{\apb@width}
\newcommand{\autoparbox}[2][c]{\settowidth{\apb@width}{#2}\parbox[#1]{\apb@width}{#2}}

\makeatother

\numberwithin{equation}{section}

\def\beq{\begin{equation}}
\def\eeq{\end{equation}}

\def\bea{\begin{eqnarray}}
\def\eea{\end{eqnarray}}

\def\beq{\begin{equation}}
\def\eeq{\end{equation}}
\def\bea{\begin{eqnarray}}
\def\eea{\end{eqnarray}}

\def\cR{{\cal R}}

\def\0{{\boldsymbol 0}}

\DeclareRobustCommand{\SkipTocEntry}[4]{}

\newcommand{\vev}[1]{\langle #1 \rangle}

\begin{document}

\begin{titlepage}

\setcounter{page}{1} \baselineskip=15.5pt \thispagestyle{empty}
 \begin{flushright}YITP-17-03,~IPMU17-0011\end{flushright}
\bigskip\

\vspace{1cm}
\begin{center}
    
{\fontsize{20}{28}\selectfont  \sffamily \bfseries
CMB Scale Dependent Non-Gaussianity  \\ \vspace{0.35cm}
from Massive Gravity during Inflation}
\end{center}

\vspace{0.2cm}

\begin{center}
{\fontsize{14}{30}\selectfont   Guillem Dom\`enech$^1$, 
Takashi Hiramatsu$^{1,2}$, Chunshan Lin$^1$, Misao Sasaki$^1$, \\
Maresuke Shiraishi$^3$, Yi Wang$^4$}
\end{center}

\begin{center}
\textsl{$^1$ Center for Gravitational Physics,
Yukawa Institute for Theoretical Physics, Kyoto University, 606-8502, Japan}
\end{center}
\begin{center}
\textsl{$^2$ Department of Physics, Rikkyo University, Toshima, Tokyo, 171-8501, Japan}
\end{center}
\begin{center}
\textsl{$^3$Kavli Institute for the Physics and Mathematics of the Universe (Kavli IPMU, WPI), \\
UTIAS, The University of Tokyo, Chiba, 277-8583, Japan}
\end{center}
\begin{center}
\textsl{$^4$Department of Physics, The Hong Kong University of Science and Technology, \\
Clear Water Bay, Kowloon, Hong Kong, P.R.China}
\end{center}

\vspace{1.2cm}
\hrule \vspace{0.3cm}
\noindent {\sffamily \bfseries Abstract} \\[0.1cm]
We consider a cosmological model in which the tensor 
mode becomes massive during inflation, and study 
the Cosmic Microwave Background (CMB) temperature 
and polarization bispectra arising from the mixing between 
the scalar mode and the massive tensor mode during inflation. 
The model assumes the existence of a preferred spatial frame during inflation.
The local Lorentz invariance is already broken in cosmology due to the
existence of a preferred rest frame. The existence of a
preferred spatial frame further breaks the remaining local SO(3)
invariance and in particular gives rise to a mass in the tensor mode. 
At linear perturbation level, we minimize our model so that the vector 
mode remains non-dynamical, while the scalar mode is the same as 
the one in single-field slow-roll inflation. 
At non-linear perturbation level, this inflationary massive graviton phase 
leads to a sizeable scalar-scalar-tensor coupling, much greater than 
the scalar-scalar-scalar one, as opposed to the conventional case. 
This scalar-scalar-tensor interaction imprints a scale dependent 
feature in the CMB temperature and polarization bispectra.
Very intriguingly, we find a surprizing similarity between
the predicted scale dependence and the scale-dependent non-Gaussianities 
at low multipoles hinted in the WMAP and Planck results.
\
\vskip 10pt
\hrule

\vspace{0.6cm}
 \end{titlepage}

 \tableofcontents

\newpage

\section{Introduction}
Inflation is the leading paradigm of the very early universe cosmology~\cite{inf1,inf2,inf3,inf4}.
It successfully explains the CMB temperature fluctuation and large scale structure 
that we observe nowadays. Despite the great success of inflationary paradigm, 
its origin still remains unknown. One of the simplest theoretical possibilities 
is that inflation was driven by a single scalar field, called inflaton,
 rolling on a slightly tilted platform of its potential. In the regime
where the inflaton moves sufficiently slowly and monotonically, 
there is a one-to-one correspondence between the value of the inflaton
and the cosmic time, giving rise to a preferred time slicing defined
by the dynamics of the inflaton.

Recently, there has been a growing interest in the inflationary physics
 with broken spatial 
reparameterisation \cite{Lin:2015cqa,Endlich:2012pz,Bartolo:2013msa,Akhshik:2014gja,Cannone:2014uqa,Hidaka:2014fra,Lin:2015nda,Cannone:2015rra}.
 From the general relativistic point of view, one introduces
3 scalar fields which determine a preferred spatial frame. 
The configuration of the 3 scalar fields is such that 
the energy momentum tensor is homogeneous and isotropic on the background 
but ceases to be so at the perturbation level.
This induces an effective mass to the transverse and traceless part of the 
spatial metric, i.e. the tensor mode (which is eventually identified with the
graviton). 
In particle physics language, there appear 3 Nambu-Goldstone bosons associated 
with the broken spatial re-parameterisation invariance (4 in total if we 
include the one associated with broken time re-parameterisation invariance
due to the inflaton dynamics). 
In the unitary gauge, these 3 Goldstone bosons are eaten by 
one of the scalar degrees and two degrees of the tensor mode in the
spatial metric. Thus in particular the tensor mode becomes massive.
In this sense, general 
relativity is extended to a theory with non-vanishing graviton mass. 

Searching for a finite range gravity is a basic question of classical 
field theory. It can be traced all the way back to the pioneering work 
by Fierz and Pauli in 1939 \cite{Fierz1939}. 
See Ref. \cite{ArkaniHamed:2002sp,deRham:2010kj,Hassan:2011hr,Gumrukcuoglu:2011ew,Gumrukcuoglu:2011zh,DeFelice:2013awa,Dubovsky:2004sg,Lin:2013aha,Lin:2013sja,Comelli:2014xga,DeFelice:2015hla} for some relevant references and Ref. \cite{deRham:2014zqa} 
for a comprehensive review on the recent progresses in this topic.  
Besides the motivation from a purely theoretical interest, 
the question whether gravity needs to be modified in the
very early universe is particularly
 important due to the Lyth bound~\cite{Lyth:1996im}. 
 According to it, the tensor-to-scalar ratio $r$ is proportional to 
the variation of the inflaton field during inflation.
It can be quite generically shown that
the amplitude of the primordial gravitational waves 
with $r\gtrsim 10^{-3}$ requires a super-Planckian excursion of the inflaton.
which is generally regarded as being out of the validity of the 
low energy effective field theory.  
In the near future, several next-generation satellite missions  
as well as the ground based and balloons experiments, are aimed at 
measuring primordial gravitational 
waves down to $r\sim 10^{-3}$ \cite{Bouchet:2011ck,Hazumi:2012aa,Abazajian:2016yjj}. 
Thus, the detection of a primordial tensor perturbation with sufficiently large 
amplitude has a profound impact on our understanding of fundamental physics. 
It implies either gravity or quantum field theory needs to be modified 
in the very early universe. See \cite{Lin:2015nda} for example for
a mechanism to evade the Lyth bound by means of a parametric resonance 
in the context of massive gravity.


Another key focus in modern cosmology is non-Gaussianities 
in the CMB fluctuations. WMAP reported a hint of a local-type
 non-Gaussianity, but the confidence level was smaller than $2\sigma$ and 
thus nothing firm was concluded \cite{Komatsu:2010fb,Bennett:2012zja}. 
The Planck satellite improved the constraints on non-Gaussianities
with high multipole $\ell$ data with better precision \cite{Ade:2013ydc,Ade:2015ava}.
The current constraint on the scale independent local-type 
non-Gaussianity is $f_{NL}^{local}=2.5\pm 5.7$ \cite{Ade:2015ava}. 
Interestingly enough, if the maximum multipole moment is taken to be 
comparable to that of WMAP (around $\ell_\mathrm{max}=500$), 
the Planck result also indicates a positive 
local non-Gaussianity, while for $\ell_\mathrm{max}>1000$, 
$f^{local}_{NL}$ found to be vanishingly small.
This suggests a scale dependent local non-Gaussianity. The bispectrum 
for the E-mode polarization shows a very similar pattern as well, 
providing an additional support for such scale dependence. 

Assuming that the scale dependent local non-Gaussianity is
of primordial origin, there can be roughly two straightforward explanations. 
The first possibility is to tune the scale dependence 
by controlling the parameters in multi-field 
inflation~\cite{Linde:1996gt, Sasaki:2006kq, Dvali:2003em, Kofman:2003nx, Suyama:2007bg}
or quasi-single field inflation \cite{Chen:2009we, Chen:2009zp, Baumann:2011nk}
(where quasi-local-type non-Gaussianity can be generated).
The first constraint on the running of non-Gaussianity of scalar perturbations can be found in the Ref.~\cite{Becker:2012je}.
The second possibility is to generate the scale dependence
through the scalar-scalar-tensor coupling. 
However, the tensor mode produces a much smaller amplitude compared 
to that induced by the scalar mode (recall that the 
tensor-to-scalar ratio is less than $0.1$). 
Nevertheless, if the tensor mode were highly non-Gaussian,
they could significantly contribute to non-Gaussianities 
in the temperature and E-mode polarization maps.
Some similar ideas have been pursued to generate low $\ell$ CMB anomalies
such as power asymmetries \cite{Chen:2014eua} 
and quadrupolar anisotropy \cite{Bordin:2016ruc}.
Resorting to the tensor mode for a scale-dependent non-Gaussianity
has an advantage, compared to the multi-field or quasi-single field case,
that one only needs to fit the amplitude since the scale dependence 
is naturally obtained by the decay of the tensor perturbations 
(gravitational waves) after they enter the horizon.

In this paper, we present a novel model where 
a sufficiently large scale-dependent local-type non-Gaussianity 
is generated from the scalar-scalar-tensor coupling.
This has not been realized before,
 to the best of our knowledge, simply due to the fact that, since $r<0.1$,
 all the inflationary models in the literature which generates 
a large tensor non-Gaussianity generates an even larger scalar 
non-Gaussianity.%
\footnote{See Ref.~\cite{Namba:2015gja} for an inflationary model 
with an axion-like spectator field realizing an equilateral-type 
non-Gaussianity sourced mainly by the gravitational wave sector.}
As a result, the scale-dependence of the non-Gaussianity due to 
the tensor mode is swamped by the scale-independent non-Gaussianity 
from the scalar sector, and is constrained to be small by the Planck 
result on non-Gaussianities up to $\ell\sim 2000$.

In the current work, we show that with the help of spatial symmetry 
breaking, the tensor non-Gaussianity can be much larger than the scalar 
non-Gaussianity. The paper is organized as follows. 
In Section \ref{sec:ssbdi}, we build an inflationary model
with spatial symmetry breaking. In Section \ref{sec:pps}, we calculate 
the primordial fluctuations of the model, including the scalar and
tensor power spectra and the non-Gaussianities. 
In Section \ref{sec:ctb}, we compute the 
CMB temperature and polarization bispectra from the primordial 
scalar-scalar-tensor coupling. We conclude in Section \ref{sec:cd}.
 Further details on the calculations are presented in Appendix.

\section{Spacetime Symmetry Breaking during Inflation}
\label{sec:ssbdi}
 
In this section, we build our model by means of 
the low energy Effective Field Theory approach.
We consider a spatially flat background with the 3-metric of the form,
\begin{equation}
ds_{(3)}^2=a^2\delta_{ab}e^{(a)}_ie^{(b)}_jdx^idx^j\,,
\end{equation}
where $a$ is the cosmic scale factor, and $e^{(a)}_i$ is
the triad basis representing the local SO(3) symmetry,
which satisfies $\delta_{ab}e^{(a)}_ie^{(b)}_j=\delta_{ij}$,
with $a$, $b$ being the internal indices and $i$, $j$ the space indices, 
On this background we consider
a model in which there is a preferred spatial frame during inflation. 
Namely the local SO(3) symmetry is broken by the existence of 
a preferred, rigid spatial frame, say $e^{(a)}_i=\delta^a_i$.

A minimal description for this preferred spatial frame is to 
introduce 3 Stueckelberg scalar fields,
\begin{eqnarray}
\varphi^a=\delta^a_ix^i+\pi^a,
\end{eqnarray}
where $\pi's$ are three Nambu-Goldstone bosons that nonlinearly 
recover the local SO(3) symmetry.
 The internal symmetries we impose on our theory 
are the SO(3) rotational symmetry and rescaling symmetry,
\begin{eqnarray}\label{ressym}
\varphi^a\to\Lambda^a{}_b\varphi^b,\hspace{10mm}\varphi^{a}\to \lambda\varphi^{a}.
\end{eqnarray}
In passing we mention that since there always exists a preferred time 
slicing or rest frame in cosmology, and the inflaton field $\phi$ determines
the preferred rest frame during inflation, one may regard
the set $(\phi,\varphi^a)$ as the 4 Stueckelberg scalar fields 
that recovers the local Lorentz (i.e. SO(3,1)) symmetry.
However, below we will not take this view but only focus on
the spatial symmetry breaking.

We follow the useful notation introduced in Ref.~\cite{Lin:2015cqa},
\begin{eqnarray}\label{eq:Z}
Z^{ab}\equiv g^{\mu\nu}\partial_{\mu}\varphi^a\partial_{\nu}\varphi^b,
\hspace{10mm}\bar{\delta}Z^{ab}\equiv \frac{Z^{ab}}{Z}
-3\frac{\delta_{cd}Z^{ac}Z^{bd}}{Z^2},
\end{eqnarray}
where $Z$ is the trace of $Z^{ab}$. 
These nontrivial VEVs of the fields are the origin of a non-vanishing graviton
mass. The action can be presumptively written as 
\begin{eqnarray}\label{act}
S=\int d^4x\sqrt{-g}\left[\frac{1}{2}M_p^2\mathcal{R}
-\frac{1}{2}g^{\mu\nu}\partial_{\mu}\phi\partial_{\nu}\phi
-V(\phi)-\frac{9}{8}M_p^2m_g^2(\bar{\delta}Z^{ab})^2+\cdots\right],
\end{eqnarray}
where $m_g^2$ is the graviton mass, 
$(\bar{\delta}Z^{ab})^2=\delta_{ac}\delta_{bd}\bar{\delta}Z^{ab}\bar{\delta}Z^{cd}$,
and the last dots stand for the higher order operators which 
only appear at nonlinear perturbation level. We assume a functional dependence of $m_g^2$ on another scalar field, e.g. inflaton, so that during inflation the graviton mass scale is around the typical scale of inflation, and after inflation the graviton mass vanishes or reduce to a very small value which is below current observational bound \cite{deRham:2016nuf}. Note that $\bar{\delta}Z^{ab}$ is defined in the way that it does not 
contribute to the background energy momentum tensor, 
and it is  traceless  at linear perturbation level. 
Thus, the background dynamics is totally determined by the 
slow rolling of the inflaton scalar field. 

According to the Nambu-Goldstone theorem, one may also expect 
3 additional Nambu-Goldstone bosons to appear in the inflationary
 perturbation spectra, in addition to the scalar degree associated 
with the inflaton scalar. In the unitary gauge, in which we set $\pi^a=0$,
one would also expect these 3 Nambu-Goldstone bosons are eaten by 
the degrees of freedom in the spatial metric, which we call the graviton
for simplicity, and the graviton develops one helicity 0 and two helicity 1 modes. 
However, this is not always necessarily the case due to the 
Lorentz symmetry breaking (i.e. the existence of the preferred rest
frame) of the background.
In fact these 3 degrees of $\pi's$ turn out to be non-dynamical
at leading order in gradient expansion, in other words,
at tree level in the context of quantum field theory.  

The reason is because of the $SO(3)$ rotational symmetry and
the rescaling symmetry given by Eq.~(\ref{ressym}).
Consider a perturbation $\pi^a$ in the long wavelength limit.
In this limit, the general form of $\pi^a$ will be
\begin{align}
\pi^a=\rho(x)\delta^{a}_jx^j+\omega^{ab}(x)\delta_{bj}x^j\,,
\label{longlim}
\end{align}
where $\omega^{ab}$ is antisymmetric in its indices
and the spatial derivatives of $\rho$ and $\omega^{ab}$ are assumed
to negligible in comparison with the Hubble scale, namely,
\begin{align}
\partial_i\rho\ll H\rho\,,
\quad
\partial_i\omega^{ab}\ll H\omega^{ab}\,.
\end{align}
The term proportional to $\rho$ in Eq.~(\ref{longlim}) 
corresponds to the helicity 0 mode (scalar-type perturbation), 
and the one proportional to $\omega^{ab}$ to the helicity 1 modes
(vector-type perturbation).
It is then easy to see that these two types of perturbations just
reduce to an infinitesimal symmetry transformation~(\ref{ressym})
 at leading order in gradient expansion. Hence they cannot be dynamical.

In what follows, for the sake of notational simplicity,
 we identify the internal
indices $a,b,\cdots$ with the space indices $i,j,\cdots$, which
actually corresponds to the particular choice of the spatial frame
$e^{(a)}_i=\delta^a_i$,
and also omit the kronecker delta in summation over the indices,
e.g. $\delta_{ij}\pi^i\pi^j\to \pi^i\pi^i$, unless there is
a chance of confusion.

In the decoupling limit, 
the quadratic action of Nambu-Goldstone bosons are calculated as
\begin{eqnarray}
S_{\pi}&=&\frac{9}{4}\Lambda^4\int \partial_i\pi^j\partial_i\pi^j
+\frac{1}{3}\left(\partial_i\pi^a\right)^2\nonumber\\
&\sim&\Lambda^4 k^2\pi^i\pi^i,
\end{eqnarray}
where $\Lambda^2\equiv M_pm_g$ is the UV cut-off of our theory. 
The absence of the kinetic term at the level of lowest dimensional operator 
implies that these 3 Nambu-Goldstone bosons require the kinetic term 
from higher dimensional operators. For instance, if we include the term
$g^{\mu\nu}\partial_{\mu}\bar{\delta}Z^{ij}\partial_{\nu}\bar{\delta}Z^{ij}$, 
the Nambu-Goldstone boson action schematically reads
\begin{eqnarray}
S_{\pi}=\Lambda^4k^2(\pi^i)^2+\Lambda^2k^2(\dot{\pi}^i)^2.
\end{eqnarray}
After canonical normalisation $\pi^i\to\Lambda k\pi^i$, we found 
these 3 Nambu-Goldstone bosons become supermassive $m^2_{\pi}\sim \Lambda^2$.
 During inflation, the energy scale of the mixing between the inflaton 
and gravity is characterised by $\Lambda_{mix}^2\sim\dot{H}$.
 In a massive gauge field theory, the mixing scale is generally characterised
 by the mass scale of the gauge field. If we assume that the massive gravity and
 inflation are governed by the same physics, it is natural to identify 
the graviton mass with the mixing scale, namely we have $m_g^2\sim\dot{H}$.
 Then the mass of these 3 Nambu-Goldstone bosons will be given by
$m_\pi^2\sim M_p\dot{H}^{1/2}\gg H^2$. Thus any excitations
of these bosons exponentially decay away during inflation,
and their existence becomes completely ignorable. 

Assuming the above argument is more or less valid, 
we will simply neglect these 3 Nambu-Goldstone pions.
Then considering these 3 modes in the unitary gauge, it is apparent
that the scalar and vector type perturbations corresponding to these
3 modes remain non-dynamical. Namely, the momentum constraints still
kill the 3 degrees of freedom.
Meanwhile, as we will see in the section \ref{sec:LP},
 the tensor mode (gravitational waves) receives a modest mass correction 
$m_{GW}^2\sim m_g^2\ll H^2$. The broken Lorentz invariance, as well as 
the remaining symmetry of the scalar field configuration (\ref{ressym})
is the origin of this mass hierarchy. 

Therefore, at low energy scale, our model has only 3 degrees of freedom;
one scalar mode associated with the fluctuations of the inflaton 
field and two massive tensor modes. We will confirm this point in the 
following detailed perturbation analysis.  In the long wavelength 
limit $k\to 0$, one may worry that the kinetic term vanishes again and 
our theory becomes strongly coupled. However, a small but non-vanishing 
kinetic term of Nambu-Goldstone pion can always arise from loop corrections 
or global symmetry $\varphi^i\to\lambda\varphi^i$ breaking at ultra-low 
energy scale, and protects our theory from the strong coupling problem. 

In principle, there are also possible problematic terms such as
 $(\partial^2\pi^a)^2$ at the level of higher order derivatives. 
The health of our theory requires that
these terms should appear in the Galileon form \cite{Nicolis:2008in},
\begin{eqnarray}
\nabla^2\pi^a\nabla^2\pi^b
-(\nabla_{\mu}\nabla_{\nu}\pi^a)(\nabla^{\mu}\nabla^{\nu}\pi^b),
\end{eqnarray}
and thus the higher order time derivative terms 
$\ddot{\pi}^a\ddot{\pi}^b$ are cancelled out to render our theory stable 
against ghost instabilities arising from higher order derivatives.

\section{Primordial Perturbation Spectra}
\label{sec:pps}
\subsection{Linear Perturbation}\label{sec:LP}
At the level of lowest dimensional operators, the 3 Goldstone bosons are non-dynamical and can be integrated out. After that, the only dynamical degrees are two tensor modes and one scalar mode from inflaton. We use ADM formalism to decompose metric,
\begin{eqnarray}
ds^2=-N^2dt^2+h_{ij}\left(N^idt+dx^i\right)\left(N^jdt+dx^j\right),
\end{eqnarray}
where, 
neglecting vector perturbations since they are non-dynamical as we have pointed out in the previous section, one has
\begin{eqnarray}
N=1+\alpha,\hspace{10mm}
N_i=\partial_i\beta,\hspace{10mm}h_{ij}
=a^2e^{2\cR}\exp\left(\gamma_{ij}+\partial_i\partial_jE\right).
\end{eqnarray}
where $\alpha,~\beta,~\cR,~E$ are scalar metric perturbation, and 
$\gamma_{ij}$ is the tensor perturbation which satisfies transverse and 
traceless condition $\gamma^{i}{}_{i}=\partial_i\gamma^i{}_{j}=0$. 
On the other hand, we decompose the scalar fields as
\begin{align}
\phi=\phi_0+\delta\phi \quad {\rm and} \quad \varphi^i=x^i+\partial^i\pi\,,
\end{align}
where again we only focused on the scalar component. From now on we work 
in the uniform $\phi$ slicing, i.e. we set $\delta\phi=0$, and
 therefore $\cR$ corresponds to the comoving curvature perturbation.
It should be noted that due to the background $SO(3)$ symmetry, scalar 
modes and tensor modes completely decouple at linear perturbation level.
\\

\noindent{\bf Scalar-type perturbation:} 
After solving the constraints for $\alpha$ and $\beta$ at 
linear perturbation level, which yields
\begin{eqnarray}\label{cons}
\alpha=\frac{\dot{\cR}}{H}\,,\hspace{20mm}\beta
=-\frac{\cR}{H}-\frac{a^2\epsilon\dot{\cR}}{k^2}+\frac{1}{2}a^2\dot{E}\,,
\end{eqnarray}
the scalar action reads 
\begin{eqnarray}\label{s2}
S_{(2)}=M_p^2\int dt\,d^3x a^3\epsilon
\left(\dot{\cR}^2-a^{-2}(\partial_i\cR)^2\right)
-\frac{1}{12}m_g^2k^4a^3(E-\pi)^2\,,
\end{eqnarray}
where $\epsilon=-\dot H/H^2$ is the slow-roll parameter. 
The graviton mass term gives us an addition constraint.
For $\pi=0$, i.e. in the so-called unitary gauge, we obtain $E=0$ by 
solving the constraint\footnote{With higher order derivative term such 
as $\nabla^{\mu}\bar{\delta}Z^{ij}\partial_{\mu}\bar{\delta}Z^{ij}$ included,
the mode $E$ becomes dynamical again but with mass $m_E^2\sim M_pm_g\gg H^2$. 
It decays exponentially on the inflationary background and thus we can neglect
 its contribution to the perturbation spectra. This is also the case 
for the vector perturbations.}. After integrating out the $E$ component,
 Eq. (\ref{s2}) can be rewritten in terms of the conformal time, 
\begin{eqnarray}
S_{(2)}=\frac{1}{2}\int d\tau\,d^3x\, z^2
\left({\cR'}^2-(\partial_i\cR)^2\right)
\,;\quad z^2\equiv 2M_p^2a^2\epsilon\,,
\end{eqnarray}
where the prime $'$ denotes the derivative with respect to the conformal time
$d\tau=dt/a(t)$.

In canonical quantization, we write
\begin{eqnarray}
\cR=\int\frac{d^3k}{(2\pi)^{3/2}}
\left(a_{\bm{k}}{u}_{k}(\tau)e^{i\bm{k}\cdot{\bm{x}}}+h.c.\right),
\end{eqnarray}
where  
  ${u}_k(\tau)$ is the positive frequency mode function which
satisfies the Klein-Gordon normalization,
\begin{align}
u_ku_k^*{}'-u_k^*u_k'=\frac{i}{z^2}\,.
\end{align}
The equation of motion reads
\begin{eqnarray}\label{eq:curvature}
{u}_k''+\frac{(z^2)'}{z^2}u_k'+k^2{u}_k=0\,.
\end{eqnarray}
In the de-Sitter approximation where $a=-1/H\tau$ and $\epsilon=const.$,
which is valid for $k\gg Ha$, we have $(z^2)'/z^2=-2/\tau$.
In the high frequency limit the positive frequency function is given by
\begin{align}
u_k\to \frac{1}{z\sqrt{2k}}e^{-ik\tau}\,.
\end{align}
Then we obtain the solution in the de-Sitter approximation as
\begin{eqnarray}
u_k=\frac{H}{M_p\sqrt{4\epsilon k^3}}e^{-ik\tau}\left(1+ik\tau\right)\,,
\end{eqnarray}
where we have ignored an irrelevant overall phase.
For $k\ll Ha$, the de Sitter approximation breaks down, 
but it is easy to see that $u_k$ approaches a constant,
\begin{align}
u_k\to\frac{H_k}{M_p\sqrt{4\epsilon_k k^3}}e^{-ik\theta_k}\,,
\end{align}
where $H_k$ and $\epsilon_k$ are those evaluated at horizon crossing
$-k\tau=1$ and $\theta_k$ is a constant phase.
We define the Fourier space $\cR$ by
\begin{align}
\cR(\bm{k},\tau)=\int d^3x\cR(\bm{x},\tau) e^{-i\bm{k}\cdot\bm{x}} \,,
\end{align}

hence
\begin{align}
\cR(\bm{k},\tau)=(2\pi)^{3/2}
\left(a_{\bm{k}}{u}_{k}(\tau)+a_{-\bm{k}}^{\dagger}{u}^*_k(\tau)\right).
\end{align}

From now on, we work on the Fourier space, and use the same symbol
$\cR$ for the Fourier mode unless there is a chance of confusion.
The two-point function for $\cR$ is given by
\begin{eqnarray}
\langle \cR(\bm{k}_1,\tau_1)\cR(\bm{k}_2,\tau_2)
\rangle=(2\pi)^3\delta(\bm{k}_1+\bm{k}_2)G_{k_1}(\tau_1,\tau_2)\,,
\end{eqnarray}
where
\begin{align}
G_{k}(\tau_1,\tau_2)={u}_{k}(\tau_1){u}_{k}^*(\tau_2)\,.
\end{align}
\\

\noindent{\bf Tensor-type perturbation:} 
In contrast to General Relativity (GR), the tensor modes receive 
a mass correction due to the broken spatial symmetry,
\begin{eqnarray}
S_T^{(2)}=\frac{M_p^2}{8}\int dt d^3x a^3
\left[\dot{\gamma}_{ij}\dot{\gamma}^{ij}
-\left(\frac{k^2}{a^2}+m_g^2\right)\gamma_{ij}\gamma^{ij}\right]\,,
\label{Taction}
\end{eqnarray}
or in terms of the conformal time,
\begin{eqnarray}
S_T^{(2)}=\frac{M_p^2}{8}\int d\tau\,d^3x\,a^2
\left[{\gamma}_{ij}'{\gamma}^{ij}{}'
-\left({k^2}+m_g^2a^2\right)\gamma_{ij}\gamma^{ij}\right]\,.
\end{eqnarray}
Noted that we have neglected the contributions from higher order derivatives 
discussed in the previous section since they are small. 
As before we decompose the field into its Fourier modes,
\begin{eqnarray}\label{eq:tensormodes}
\gamma_{ij}(x)= \int \frac{d^3k}{(2\pi)^3}
\sum_{s=\pm}\epsilon^s_{ij}(\bm{k})\gamma^s(\bm{k},\tau)e^{i\bm{k}\cdot\bm{x}}\,;
\quad
\gamma^s(\bm{k},\tau)
=(2\pi)^{3/2}\left[b^s_{\bm{k}}\gamma_{k}+b^s_{-\bm{k}}{}^\dag\gamma_k^*\right]\,,
\end{eqnarray}
where $b^s_{\bm{k}}$ is the annihilation operator and $s$ is the polarisation index. 
The polarization tensor $\epsilon^s_{ij}$ obeys
\begin{align}
\epsilon_{ii}^{s}(\bm{k}) = {k}^i \epsilon_{ij}^{s}(\bm{k}) = 0\,,
\quad 
\epsilon_{ij}^{s *}(\bm{k}) = \epsilon_{ij}^{-s}(\bm{k}) 
= \epsilon_{ij}^{s}(- \bm{k})\,,
\quad
\epsilon_{ij}^{s}(\bm{k}) \epsilon_{ij}^{s'}(\bm{k}) = 2\delta_{s, -s'}\,.
\label{poltensor}
\end{align}
The action (\ref{Taction}) yields the equation of motion
\begin{align}\label{eomgamma}
{\gamma}_k''+\frac{(a^2)'}{a^2}\gamma_k'+\left(k^2+m_g^2a^2\right){\gamma}_k=0\,,
\end{align}
where again the positive frequency mode function satisfies
the Klein-Gordon normalizaion,
\begin{align}
\gamma_k\gamma_k^*{}'-\gamma_k^*\gamma_k'=\frac{i}{2a^2}\,,
\end{align}
where the factor 2 in the denominator is due to the factor 2 in
the normalization of the polarization tensor defined in Eq.~(\ref{poltensor}).

Below we assume $m_g^2\ll H^2$ (although interesting physics may arise 
when $m_g^2\gtrsim H^2$, similar to the case studied in
\cite{Chen:2012ge, Pi:2012gf, Gong:2013sma, Chen:2015lza}). 
Assuming the Bunch-Davies vacuum initial condition,
the solution to Eq. (\ref{eomgamma}) under the de Sitter approximation
is given by a Hankel function of the first kind,
\begin{eqnarray}\label{exsol}
{\gamma}_k= \frac{H}{M_p}
\sqrt{\frac{\pi}{2k^3}}(-k\tau)^{3/2}H_{\nu_g}^{(1)}\left(-k\tau\right)\,,
\end{eqnarray}
where $\nu_g=\sqrt{9/4-m_g^2/H^2}$.
In the massless limit $m_g^2\to0$, the solution reduces to 
\begin{eqnarray}\label{eq:subhorizon}
\gamma_k=\frac{iH}{M_p\sqrt{k^3}}e^{-ik\tau}\left(1+ik\tau\right).
\end{eqnarray}
In the case of a small but non-vanishing graviton mass, $m_g^2/H^2\ll1$,
we find on superhorizon scales
\begin{eqnarray}\label{eq:suphorizon}
\gamma_k\simeq\frac{H_k}{M_p\sqrt{k^3}}\left(-k\tau\right)^{m_g^2/3H^2}\,.
\end{eqnarray}
 Note that, as opposed to GR, the tensor modes are no longer constant 
on superhorizon scales. It gains a tiny time dependence. 
The larger the graviton mass is, the faster it decays.
One can easily see this behavior by solving Eq.~\eqref{eomgamma} 
in the limit $k^2\to0$ in terms of the number of e-folds, 
that is $dN=aHd\tau$. The power spectrum of primordial gravitational waves
 is thus calculated as 
\begin{eqnarray}
P_{\gamma}=\frac{2H^2}{\pi^2M_p^2}\left(\frac{k}{aH}\right)^{n_t},
\end{eqnarray}
with a tilt
\begin{eqnarray}
n_t\simeq -2\epsilon +\frac{2m_g^2}{3H^2}.
\end{eqnarray}
Interestingly if $m_g^2>3H^2\epsilon$, the primordial gravitational
 waves have a blue tilt, which could be a smoking gun of massive gravity
 in the early universe (for other possibilities of blue tilted 
tensor modes, see \cite{Wang:2014kqa}). Similar to the scalar mode,
 it is useful to define the tensor two-point function 
\begin{eqnarray}
\langle \gamma^s(\bm{k}_1,\tau_1)\gamma^{\tilde{s}}(\bm{k}_2,\tau_2)\rangle
=(2\pi)^3\delta(\bm{k}_1+\bm{k}_2)W_{k_1}(\tau_1,\tau_2)\delta_{s\tilde{s}},
\end{eqnarray}
where 
\begin{align}
W_k(\tau_1,\tau_2)=\gamma_k(\tau_1)\gamma_k^*(\tau_s)\,.
\end{align}
It should be noted that, under the small mass approximation 
$m_g^2\ll H^2$, as in the single field case, 
the tensor to scalar ratio is given by 
\begin{eqnarray}\label{r}
r\equiv\lim_{-k\tau\ll1}
\frac{2\times 2W_k(\tau,\tau)}{G_k(\tau,\tau)}\sim 16\epsilon\,,
\end{eqnarray}
where the additional factor 2 on the right-hand side is again
due to our normalization of the polarization tensor (\ref{poltensor}).

\subsection{Non-linear Perturbations}

At cubic action level, the graviton mass term in the action eq. (\ref{act}) reads
\begin{eqnarray}\label{eq:s3}
S_g&=&-\frac{9}{8}\int d^4x\sqrt{-g}M_p^2m_g^2\bar{\delta}Z^{ij}\bar{\delta}Z^{ij}\nonumber\\
&\supset&M_p^2m_g^2\int-\frac{1}{4}a^5\gamma_{ij}N^iN^j-\frac{1}{8}a^3(\alpha+3\cR)\gamma_{ij}\gamma_{ij}+\frac{3}{8}a^3\gamma_{ij}\gamma_{jk}\gamma_{ki}\,.
\end{eqnarray}
There are three types of interactions, the first term of the above equation is scalar-scalar-tensor, the second term is scalar-tensor-tensor, and the last term is with 3 tensors. In our current work, we focus on the CMB temperature and E-mode polarisation non-Gaussianities arising from primordial gravitational waves, provided that primordial three-scalar interaction is negligible. Noted that after inflation, gravitational waves decay as $a^{-1}$ after re-entering the cosmic horizon, while scalar perturbation underwent baryon acoustic oscillations, and co-evolve with matter sector at late time. Naively we would expect that the contributions from the above three types primordial coupling scales as $a^{-1}$, $a^{-2}$ and $a^{-3}$ respectively. Therefore, on the observational scales that we are interested, we consider the primordial scalar-scalar-tensor coupling as the dominant contribution to the CMB temperature and E-mode polarisation non-Gaussianities.

Plug in the solution of the momentum constraint Eq.~(\ref{cons}), 
the leading contribution in the scalar-scalar-tensor interaction is given by
\begin{align}
S_{sst}=-\frac{1}{4}M_p^2\int a^3\frac{m_g^2}{H^2}
 \gamma_{ij}\partial_i\cR\partial_j\cR+...
\end{align}
We have learned that the non-trivial scalar-scalar-tensor correlation arises 
from the coupling term such as $\gamma_{ij}N^iN^j$ in Eq.~(\ref{eq:s3}).
 In fact, such a term also arises from the cubic coupling between 
the spacelike Stueckelberg fields and the inflaton scalar field,
\begin{eqnarray}
-\frac{1}{2}\lambda\int \sqrt{-g} \frac{\bar{\delta}Z^{ij}}{Z}
\cdot\partial^{\mu}\varphi^i\partial_{\mu}\phi
\cdot\partial^{\nu}\varphi^j\partial_{\nu}\phi=-\lambda M_p^2H^2\epsilon
 \int a^5\gamma_{ij}N^iN^j ,
\end{eqnarray} 
where $\lambda$ is a free parameter of our low energy effective field theory, 
whose origin can be understood only after we have UV completion of our theory.
Here we do not discuss this issue since it is beyond the scope of the current work. 
Including the contribution from graviton mass term, Eq.~(\ref{eq:s3}),
 the scalar-scalar-tensor coupling term takes the form,
\begin{eqnarray}
S^{(3)}\supset-\lambda_{sst}M_p^2H^2\epsilon 
\int a^3\left( \frac{1}{a^2H^2}\gamma_{ij}\partial_i\cR\partial_j\cR
 -\frac{2\epsilon}{H}\gamma_{ij}\partial_i\cR\partial_j\partial^{-2}\dot{\cR}
+a^2\epsilon^2\gamma_{ij}
\partial_i\partial^{-2}\dot{\cR}\partial_j\partial^{-2}\dot{\cR}\right),
\end{eqnarray}
where 
\begin{eqnarray}\label{lamda}
\lambda_{sst}=\frac{m_g^2}{4H^2\epsilon}+\lambda.
\end{eqnarray}

\subsection{Primordial 3-point Correlation Function}
The interaction Hamiltonian reads
\begin{eqnarray}\label{Hint}
H_{int}=-L_{int}=\lambda_{sst}M_p^2H^2\epsilon a^3\left( \frac{1}{a^2H^2}\gamma_{ij}\partial_i\cR\partial_j\cR-\frac{2\epsilon}{H} \gamma_{ij}\partial_i\cR\partial_j\partial^{-2}\dot{\cR}+a^2\epsilon^2\gamma_{ij}\partial_i\partial^{-2}\dot{\cR}\partial_j\partial^{-2}\dot{\cR}\right),
\end{eqnarray}
and the 3-point function can thus be calculated by means of the in-in formalism as 
\begin{eqnarray}\label{3p1}
& &\langle \gamma_{\bm{k}_1}^s\cR_{\bm{k}_2}\cR_{\bm{k}_3}\rangle_1\nonumber\\
&=&2!\times{\rm Re}\left[2i\lambda_{sst}M_p^2H^2\epsilon \int_{-\infty(1-i\epsilon)}^{\tau}\frac{d\tilde{\tau}}{\tilde{\tau}^2}\langle  0|\gamma_{\bm{k}_1}^s(\tau)\cR_{\bm{k}_2}(\tau)\cR_{\bm{k}_3}(\tau)\gamma_{\bm{k}_4}^{s}(\tilde{\tau})\epsilon_{ij}^{s}k_{5i}k_{6j}\cR_{\bm{k}_5}(\tilde{\tau})\cR_{\bm{k}_6}(\tilde{\tau})|0\rangle\right]_{\tau\to0}\nonumber\\
&=&2!\times{\rm Re}\left[2i\lambda_{sst}M_p^2H^2\epsilon \epsilon_{ij}^{-s}k_{2i}k_{3j}\int_{-\infty(1-i\epsilon)}^{0}\frac{d\tilde{\tau}}{\tilde{\tau}^2}W_{k_1}(0,\tilde{\tau})G_{k_2}(0,\tilde{\tau})G_{k_3}(0,\tilde{\tau})\right]\nonumber\\
&=&-\left(2\pi\right)^3\delta^{(3)}\left(\sum \bm{k}_i\right)\cdot \frac{1}{\Pi\left(k_i\right)^3}\cdot\frac{\lambda_{sst}H^4}{4\epsilon M_p^4}\cdot\left(-k_t+\frac{\sum_{i<j}k_ik_j}{k_t}+\frac{k_1k_2k_3}{k_t^2}\right)\epsilon_{ij}^{-s}k_{2i}k_{3j}.
\end{eqnarray}
where $k_t\equiv k_1+k_2+k_3$, and the subscript ``1'' of $\langle...\rangle_1$
stands for the 1st term in the parentheses on the right-hand side of Eq.~(\ref{Hint}). 
We have taken the massless approximation $m_g^2/H^2\ll 1$ to compute the above integral.   Noted that  this term is the same as the one from the cubic scalar interaction
in single-field slow-roll inflation in the context of 
Einstein gravity\cite{Maldacena:2002vr}, except for the
overall coefficient $\lambda_{sst}$. A similar term has also been spotted in an extended effective field theory of inflation \cite{Bartolo:2015qvr}. 
It may dominate over the cubic scalar contribution
and thus leave an imprint on the CMB non-Gaussianity
for a sufficiently large $\lambda_{sst}$.

The contribution from the next-to-leading order in slow-roll parameter
 (2nd term in the parentheses) reads
\begin{eqnarray}\label{3p2}
& &\langle \gamma_{\bm{k}_1}^s\cR_{\bm{k}_2}\cR_{\bm{k}_3}\rangle_2\nonumber\\
&=&{\rm Re}\left[2i\lambda_{sst}M_p^2H^2\epsilon\int_{-\infty(1-i\epsilon)}^{\tau}\frac{2\epsilon}{H\left(-H\tilde{\tau}\right)^3}d\tilde{\tau}\langle  0|\gamma_{\bm{k}_1}^s(\tau)\cR_{\bm{k}_2}(\tau)\cR_{\bm{k}_3}(\tau)\gamma_{\bm{k}_4}^{s}(\tilde{\tau})
\epsilon_{ij}^{s}k_{5i}k_{6j}k_6^{-2}\cR_{\bm{k}_5}(\tilde{\tau})\cR_{\bm{k}_6}'(\tilde{\tau})|0\rangle\right]_{\tau\to0}\nonumber\\
&=& \left(2\pi\right)^3\delta^{(3)}\left(\sum \bm{k}_i\right)\cdot\frac{1}{4}\lambda_{sst}\frac{H^4}{M_p^4}\log\left(-k_t\tau\right)\epsilon^{-s}_{ij}k_{2i}k_{3j}\left(\frac{1}{ k_1^3k_2^3k_3^2}+\frac{1}{ k_1^3k_2^2k_3^3}\right),
\end{eqnarray}
where $\log\left(-k_t\tau\right)\simeq N_{k_{t}}$ is the number
of e-foldings from the time of horizon crossing to the end of inflation. 
Typically $N_{k_{t}}\simeq 50\sim 60$ for observable
cosmological large scales.  The last term in the parentheses is suppressed by
 $\epsilon^2$ and thus irrelevant. 

Summing up Eqs. (\ref{3p1}) and (\ref{3p2}) we get 
\begin{eqnarray}\label{3pt}
&&\langle \gamma_{\bm{k}_1}^s\cR_{\bm{k}_2}\cR_{\bm{k}_3}\rangle\nonumber\\
\simeq&&\left(2\pi\right)^3\delta^{(3)}\left(\sum \bm{k}_i\right)\cdot\lambda_{sst}\frac{H^4}{M_p^4}\cdot\frac{\epsilon^{-s}_{ij}k_{2i}k_{3j}}{\Pi\left(k_i\right)^3}\cdot\left[-\frac{1}{4\epsilon}\left(-k_t+\frac{\sum_{i<j}k_ik_j}{k_t}+\frac{k_1k_2k_3}{k_t^2}\right)+\frac{N_{k_{t}}}{4}\left(k_2+k_3\right)\right].\nonumber\\
\label{eq:tss1}
\end{eqnarray} 
This contribution should be compared to that coming from the 3-scalar interaction
 in GR. Recall that such terms are given by \cite{Maldacena:2002vr,Wang:2013eqj}
\begin{eqnarray}\label{eq:3s}
S_{3s}=M_p^2\int \epsilon^2 a^3\cR \dot{\cR}^2+...\qquad\text{and}
\qquad \langle \cR_{\bm{k}_1}\cR_{\bm{k}_2}\cR_{\bm{k}_3}\rangle
\sim \frac{H^4}{\epsilon}.
\end{eqnarray}
Then the ratio between $\langle \gamma_{\bm{k}_1}^s\cR_{\bm{k}_2}\cR_{\bm{k}_3}\rangle$
 and $\langle \cR_{\bm{k}_1}\cR_{\bm{k}_2}\cR_{\bm{k}_3}\rangle$ is given by 
\begin{eqnarray}
\frac{\langle \gamma_{\bm{k}_1}^s\cR_{\bm{k}_2}\cR_{\bm{k}_3}\rangle}{\langle \cR_{\bm{k}_1}\cR_{\bm{k}_2}\cR_{\bm{k}_3}\rangle}\sim \lambda_{sst}\,,
\end{eqnarray}
where we used Eq.~\eqref{3pt}. 
Thus, the scalar-scalar-tensor contribution
dominates over the scalar-scalar-scalar contribution if $\lambda_{sst}>1$.
 Noting that the non-Gaussianity arising from 
$\langle \cR_{\bm{k}_1}\cR_{\bm{k}_2}\cR_{\bm{k}_3}\rangle$ is of
the order of the slow-roll parameter $\epsilon$, 
i.e. $f_{NL}^{sss}\sim\epsilon$, we can estimate the non-Gaussianity 
from $\langle \gamma_{\bm{k}_1}^s\cR_{\bm{k}_2}\cR_{\bm{k}_3}\rangle$ as
\begin{eqnarray}
f_{NL}^{sst}\sim \lambda_{sst}\cdot\epsilon. \label{eq:fNLsst_theory}
\end{eqnarray}
Thus $\lambda_{sst} \gtrsim {\cal O}(10^3)$ will be required for
the temperature and polarization bispectra to be observed if $\epsilon \sim10^{-3}$.
This will be quantitatively estimated in the next section where
we perform numerical computations.

\section{CMB Bispectra}
\label{sec:ctb}

We now analyze the signatures of the primordial scalar-scalar-tensor correlator \eqref{eq:tss1} in the CMB temperature and polarization bispectra. For convenience, we rewrite Eq.~\eqref{eq:tss1} as
\begin{eqnarray}
\vev{\gamma_{{\bf k}_1}^{\lambda} \cR_{{\bf k}_2} \cR_{{\bf k}_3}}
 &=&
(2 \pi)^3 \delta^{(3)}\left({\bf k}_1 + {\bf k}_2 + {\bf k}_3 \right) \epsilon_{ij}^{-\lambda}({\bf k}_1) \hat{k}_{2i} \hat{k}_{3j} F_{k_1 k_2 k_3}^{(tss)} ~, \\
F_{k_1 k_2 k_3}^{(tss)} &\equiv& 
- \frac{16 \pi^4 \lambda_{sst} \epsilon A_S^2}{k_1^2 k_2^2 k_3^2}
\left[
  \frac{I_{k_1 k_2 k_3}}{k_t} \frac{k_t}{k_1}
  - \epsilon N_{\rm tot} \frac{k_2 + k_3}{k_1} 
  \right] ~, \\
I_{k_1 k_2 k_3} &\equiv&  -k_t + \frac{k_1 k_2 + k_2 k_3 + k_3
 k_1}{k_t} + \frac{k_1 k_2 k_3}{k_t^2}~,
\end{eqnarray}
where $A_S \equiv H^2 / (8 \pi^2 \epsilon M_p^2)$, and we have assumed
 that $N_{k_t} = N_{\rm tot} = {\rm const}$ for simplicity.

At linear order, the harmonic coefficients of the CMB anisotropies,
 $a_{\ell m} = \int d^2 \hat{\bf n} X(\hat{\bf n}) Y_{\ell m}^*(\hat{\bf n})$, 
from $\cR$ and $\gamma^{\pm}$ are generally expressed 
according to \cite{Shiraishi:2010sm,Shiraishi:2010kd} as
\begin{eqnarray}
  a_{\ell m}^{(s) X} &=& 
4\pi i^{\ell} \int \frac{d^3 k}{(2\pi)^{3}}
{\cal T}_{\ell(s)}^{X}(k) \cR_{\bf k}  Y_{\ell m}^*(\hat{\bf k}) ~, \\
  a_{\ell m}^{(t) X} &=& 
4\pi i^{\ell} \int \frac{d^3 k}{(2\pi)^{3}}
{\cal T}_{\ell(t)}^{X}(k)  \sum_{\lambda = \pm} \lambda^x \gamma_{\bf k}^{\lambda} \, {}_{-2 \lambda} Y_{\ell m}^*(\hat{\bf k}) ~, 
\end{eqnarray}
where ${\cal T}_{\ell (z)}^X$ denotes the scalar-mode ($z = s$) and 
tensor-mode ($z = t$) radiation transfer functions for the temperature ($X = T$) 
and the E/B-mode polarization ($X = E/B$), and $x=0$ for $X = T, E$
and $x=1$ for $X = B$. Note that $a_{\ell m}^{(s) B} = 0$. 
From these equations, we can formulate the induced CMB scalar-scalar-tensor bispectra as
\begin{eqnarray}
  \vev{a_{\ell_1 m_1}^{(t) X_1} a_{\ell_2 m_2}^{(s) X_2} a_{\ell_3 m_3}^{(s) X_3} }
  &=& \left[ \prod_{n=1}^3 
4\pi i^{\ell_n} \int_0^\infty \frac{k_n^2 d k_n}{(2\pi)^{3}}\right]
 {\cal T}_{\ell_1 (t)}^{X_1}(k_1)
 {\cal T}_{\ell_2 (s)}^{X_2}(k_2) {\cal T}_{\ell_3 (s)}^{X_3}(k_3) 
 F_{k_1 k_2 k_3}^{(tss)} \nonumber \\
&& 
 \left[ \prod_{n=1}^3 
 \int d^2 \hat{\bf k}_n
\right] \sum_{\lambda_1 = \pm} \lambda_1^{x_1} {}_{-2\lambda_1} 
Y_{\ell_1 m_1}^*(\hat{\bf k}_1)
  Y_{\ell_2 m_2}^*(\hat{\bf k}_2)
  Y_{\ell_3 m_3}^*(\hat{\bf k}_3) \nonumber \\
  && (2 \pi)^3 \delta^{(3)}\left({\bf k}_1 + {\bf k}_2 + {\bf k}_3 \right) 
\epsilon_{ij}^{-\lambda_1}({\bf k}_1) \hat{k}_{2i} \hat{k}_{3j} .
\end{eqnarray}
Employing the harmonic-space expressions of $\delta^{(3)}\left({\bf k}_1 + {\bf k}_2 + {\bf k}_3 \right) \epsilon_{ij}^{-\lambda_1}({\bf k}_1) \hat{k}_{2i} \hat{k}_{3j}$ and the law of addition of angular momentum, the $\hat{\bf k}_{1,2,3}$ integrals in the 2nd and 3rd lines are analytically performed and expressed in terms of
the products of the Wigner $3j$ and $9j$ symbols. Since the same angular integrals are realized in the model studied in Ref.\cite{Shiraishi:2010kd} (mathematically equivalent to the case for $N_{\rm tot} = 0$), we make use of their results,
which leads us to 
\begin{align}
  \vev{a_{\ell_1 m_1}^{(t) X_1} a_{\ell_2 m_2}^{(s) X_2} a_{\ell_3 m_3}^{(s) X_3} }
  &= B^{(tss) X_1 X_2 X_3}_{\ell_1 \ell_2 \ell_3}
  \left(
  \begin{array}{ccc}
  \ell_1 & \ell_2 & \ell_3 \\
  m_1 & m_2 & m_3 
  \end{array}
 \right) ~, \\
  B^{(tss) X_1 X_2 X_3}_{\ell_1 \ell_2 \ell_3}
  &\equiv \frac{(8 \pi)^{3/2}}{3} i^{\ell_1 + \ell_2 + \ell_3}
  \sum_{L_1, L_2, L_3 \in V} 
(-1)^{\frac{L_1 + L_2 + L_3}{2}} h^{0~0~0}_{L_1 L_2 L_3} 
h^{2 0 -2}_{\ell_1 L_1 2} h^{0~0~0}_{\ell_2 L_2 1} h^{0~ 0~ 0}_{\ell_3 L_3 1} 
\left\{
  \begin{array}{ccc}
  \ell_1 & \ell_2 & \ell_3 \\
  L_1 & L_2 & L_3 \\
  2 & 1 & 1
  \end{array}
 \right\} \nonumber \\
\times&  \int_0^\infty y^2 dy  
\frac{2}{\pi}  \int_0^\infty k_1^2 d
  k_1 {\cal T}_{\ell_1 (t)}^{X_1} j_{L_1}(k_1 y)
\left[ \prod_{n=2}^3  \frac{2}{\pi}  \int_0^\infty k_n^2 d
  k_n {\cal T}_{\ell_n (s)}^{X_n} j_{L_n}(k_n y)  \right] F_{k_1 k_2 k_3}^{(tss)}~,
 \label{eq:CMB_bis_general}
\end{align} 
where $h^{s_1 s_2 s_3}_{l_1 l_2 l_3}
\equiv \sqrt{\frac{(2 l_1 + 1)(2 l_2 + 1)(2 l_3 + 1)}{4 \pi}}
\left(
  \begin{array}{ccc}
  l_1 & l_2 & l_3 \\
  s_1 & s_2 & s_3
  \end{array}
 \right)$, and the $\ell$-space domain $V$ corresponds to 
\begin{eqnarray}
L_1 =
\begin{cases}
|\ell_1 \pm 2|, \ell_1 & ({\rm for \ } X_1 = T,E ) \\
|\ell_1 \pm 1| & ({\rm for \ } X_1 = B )
\end{cases}~, \ \ 
 L_2 = |\ell_2 \pm 1|~, \ \
 L_3 = |\ell_3 \pm 1| ~.
\end{eqnarray}
Using this formula, in what follows, we numerically analyze two 
parity-even correlators: TTT and EEE, and two parity-odd ones: TTB and EEB 
with $A_S = 2.4 \times 10^{-9}$. We then employ an approximation,
 $ \frac{I_{k_1 k_2 k_3}}{k_t} \approx -0.65$, to find a factorized 
form $ F_{k_1 k_2 k_3}^{(tss)} \approx \sum_{abc} A_a(k_1) B_b(k_2) C_c(k_3)$
to reduce the computational cost.


\begin{figure}[t!]
  \begin{tabular}{cc}
    \begin{minipage}{0.5\hsize}
      \begin{center}
        $(\lambda_{sst}, \epsilon) = (10^3, 10^{-3})$
        \includegraphics[width=1\textwidth]{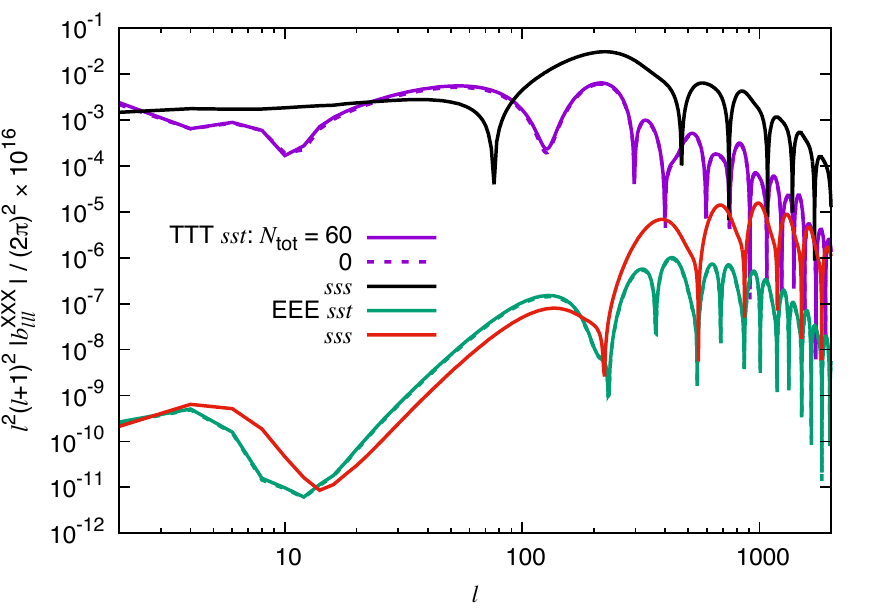}
  \end{center}
    \end{minipage}
    \begin{minipage}{0.5\hsize}
      \begin{center}
        $(\lambda_{sst}, \epsilon) = (10^2, 10^{-2})$
    \includegraphics[width=1\textwidth]{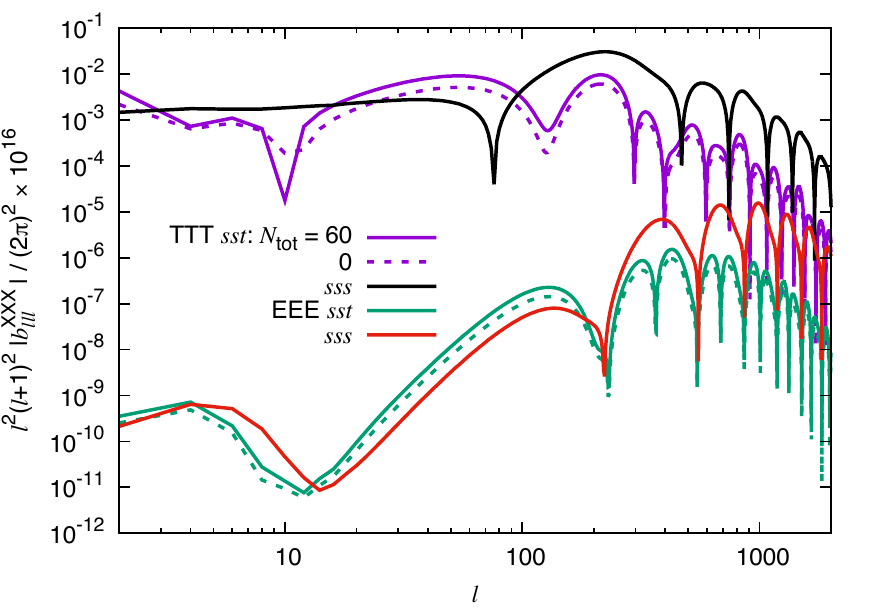}
  \end{center}
\end{minipage}
    \end{tabular}
\\
\begin{tabular}{cc}
    \begin{minipage}{0.5\hsize}
  \begin{center}
    \includegraphics[width=1\textwidth]{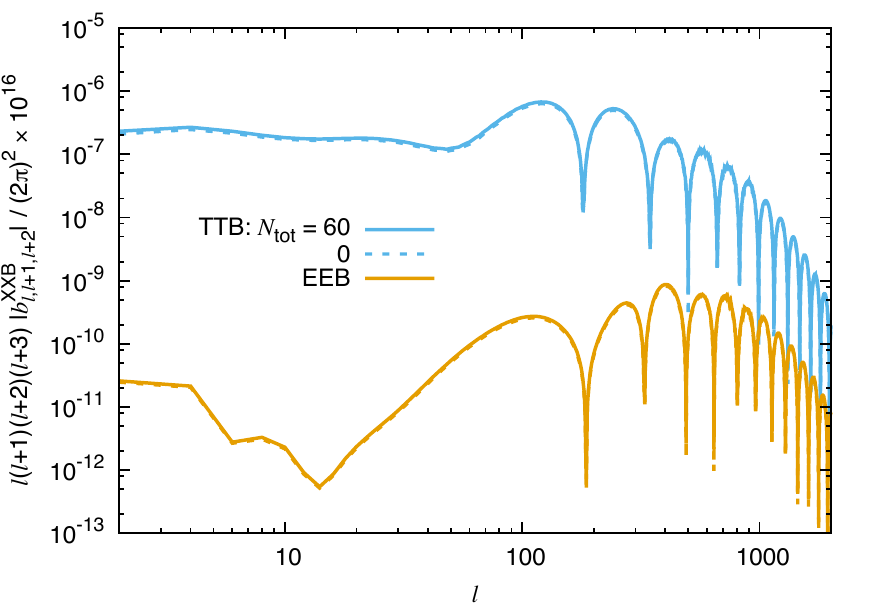}
  \end{center}
    \end{minipage}
    \begin{minipage}{0.5\hsize}
  \begin{center}
    \includegraphics[width=1\textwidth]{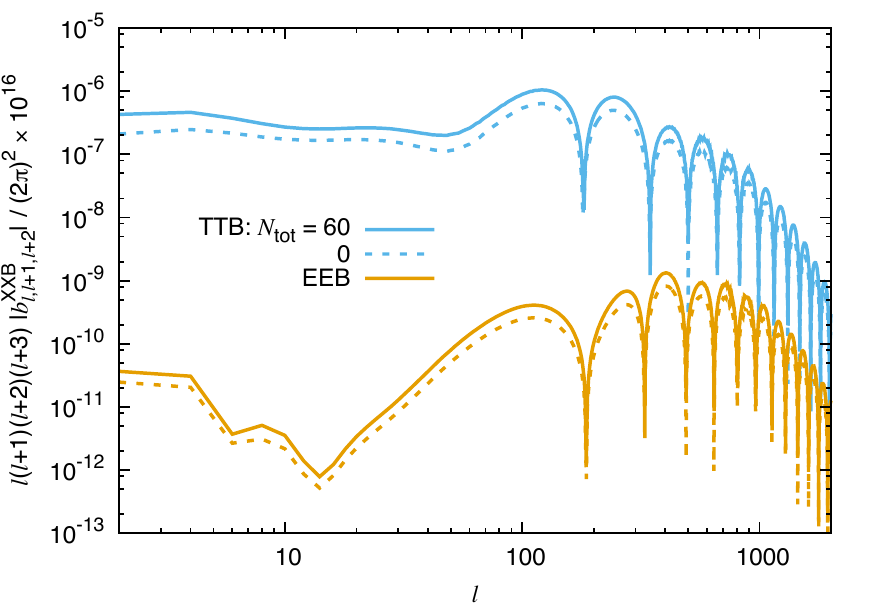}
  \end{center}
\end{minipage}
  \end{tabular}
  \caption{Equilateral-limit shapes of the TTT and EEE bispectra (top two panels), given by $b_{\ell_1 \ell_2 \ell_3}^{XXX} = [B^{(tss)XXX}_{\ell_1 \ell_2 \ell_3} + B^{(sts)XXX}_{\ell_1 \ell_2 \ell_3} + B^{(sst)XXX}_{\ell_1 \ell_2 \ell_3} ] / h^{0~0~0}_{\ell_1 \ell_2 \ell_3}$,  and the TTB and EEB bispectra (bottom two panels), given by $b_{\ell_1 \ell_2 \ell_3}^{XXB} = B^{(sst) XXB}_{\ell_1 \ell_2 \ell_3} / h^{0~0~0}_{\ell_1 +2, \ell_2 +1, \ell_3}$, from the $sst$ correlator \eqref{eq:tss1} with $(\lambda_{sst}, \epsilon) = (10^3, 10^{-3})$ and $(10^2, 10^{-2})$. Solid (dotted) line corresponds to the case for $N_{\rm tot} = 60$ ($0$). For comparison, in the top panels we also plot the TTT and EEE bispectra in the $sss$ case with $f_{\rm NL}^{sss} = 1$.}
\label{fig:blll}
\end{figure}

Figure~\ref{fig:blll} shows the equilateral-limit shapes of the TTT, EEE, TTB and EEB bispectra from the scalar-scalar-tensor correlator (referred to as $sst$) for several $(\lambda_{sst}, \epsilon, N_{\rm tot}$). Here the TTT and EEE (TTB and EEB) bispectra obey $\ell_1 + \ell_2 + \ell_3 = \text{even}$ (odd) because of no parity violation in the $sst$ correlator \eqref{eq:tss1}. One can find the expected signatures such as the amplification at $\ell \sim 200$ due to the acoustic oscillation in TTT and TTB, and the reionization bump for $\ell \lesssim 10$ in EEE and EEB. In the top panel, we also draw the TTT and EEE bispectra induced by the usual scalar-mode local-type non-Gaussianity (referred to as $sss$) with $f_{\rm NL}^{sss} = 1$ \cite{Komatsu:2001rj}. It 
can be easily seen that the $sst$ bispectra are comparable in size to the $sss$ ones up to $\ell \sim 100$, while, after that, the difference appears due to the rapid decay of the tensor-mode transfer function compared with the scalar-mode one. 
The overall amplitudes of the bispectra change depending 
on $\lambda_{sst}$, $\epsilon$ and $N_{\rm tot}$. 
We cannot, however, see the discrepancy between the bispectra 
with $N_{\rm tot} = 60$ and $0$ when $\epsilon = 10^{-3}$, 
which implies that the contribution of the 2nd term 
in Eq.~\eqref{eq:tss1} is subdominant. 


\begin{figure}[t]
 \begin{tabular}{cc}
   \begin{minipage}{0.5\hsize}
     \begin{center}
       $(\lambda_{sst}, \epsilon) = (10^3, 10^{-3})$
    \includegraphics[width=1\textwidth]{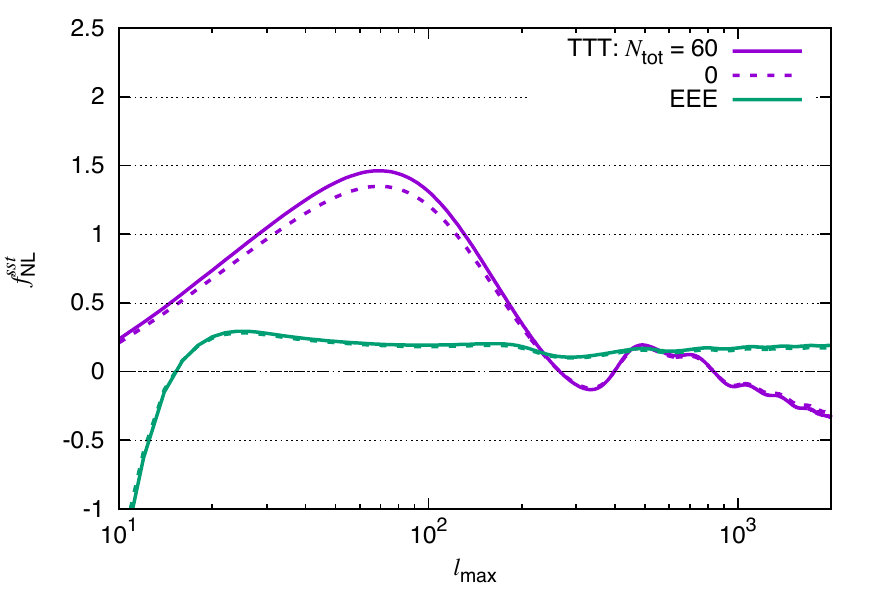}
  \end{center}
   \end{minipage}
   \begin{minipage}{0.5\hsize}
     \begin{center}
       $(\lambda_{sst}, \epsilon) = (10^2, 10^{-2})$
    \includegraphics[width=1\textwidth]{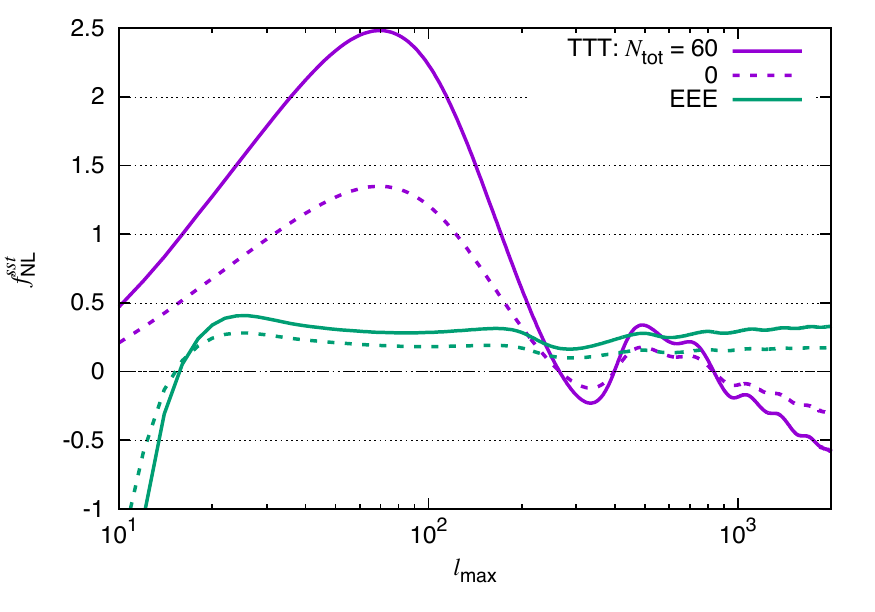}
  \end{center}
   \end{minipage}
\end{tabular}
\caption{Size of $f_{\rm NL}^{sss}$ translated from the $sst$ correlator, i.e., $f_{\rm NL}^{sst}$, for $(\lambda_{sst}, \epsilon) = (10^3, 10^{-3})$ and $(10^2, 10^{-2})$, estimated from TTT and EEE. Solid and dotted lines correspond to the cases for $N_{\rm tot} = 60$ and $0$, respectively. Given the value of $\epsilon$ fixed, the vertical axis could be understood as $f_{NL}^{sst}/(\epsilon\lambda_{sst})$ and thus $f_{NL}^{sst}$ linearly depends on $\lambda_{sst}$.}
\label{fig:fNL}
\end{figure}

To discuss more quantitatively, let us perform the Fisher matrix analysis. 
Ignoring the contributions of the higher-order correlations to the covariance 
matrix, the Fisher matrices of XXX and XXB for X$=$T or $E$ are given by
\begin{eqnarray}
  F_{ij}^{XXX} &=& \sum_{\ell_1, \ell_2, \ell_3 = 2}^{\ell_{\rm max}}
   \frac{\hat{B}_{\ell_1 \ell_2 \ell_3}^{XXX, i} \hat{B}_{\ell_1 \ell_2 \ell_3}^{XXX,j}}{6 C_{\ell_1}^{XX} C_{\ell_2}^{XX} C_{\ell_3}^{XX}} (-1)^{\ell_1 + \ell_2 + \ell_3} ~, \label{eq:fish_XXX} \\
   F_{ij}^{BXX} &=& \sum_{\ell_1, \ell_2, \ell_3 = 2}^{\ell_{\rm max}}
   \frac{\hat{B}_{\ell_1 \ell_2 \ell_3}^{BXX, i} \hat{B}_{\ell_1 \ell_2 \ell_3}^{BXX,j}}{2 C_{\ell_1}^{BB} C_{\ell_2}^{XX} C_{\ell_3}^{XX}} (-1)^{\ell_1 + \ell_2 + \ell_3} ~, \label{eq:fish_BXX}
 \end{eqnarray}
where $\hat{B}_{\ell_1 \ell_2 \ell_3}^i$ is the bispectrum in the $i$-th model normalized with the amplitude parameter, and $C_\ell$ is the power spectrum including experimental uncertainties. We consider noiseless measurements of the temperature and E-mode polarization up to $\ell = 2000$ (comparable to the sensitivity in the Planck experiment \cite{Ade:2015ava}), so $C_\ell^{TT}$ and $C_\ell^{EE}$ are determined by the cosmic variance alone. For the TTB and EEB cases, we assume the analyses with the B-mode data (without delensing) expected in a futuristic survey like the LiteBIRD project and thus $C_\ell^{BB}$ is given by the sum of the cosmic variance, the lensed B-mode spectrum and the LiteBIRD-level noise spectrum \cite{Hazumi:2012aa,Matsumura:2013aja,Matsumura:2016sri}.%
\footnote{The comparable results will be obtained even if one analyzes with the noise spectra expected in other futuristic B-mode surveys such as CMBS4 \cite{Abazajian:2016yjj} and COrE \cite{Bouchet:2011ck}.}
Since we are working with the massless approximation $m_g^2 / H^2 \ll 1$, the cosmic variance is given by the BB power spectrum with the tensor-to-scalar ratio $r \simeq 16 \epsilon$. We then consider the full-sky measurements and ignore the contaminations due to residual foregrounds for simplicity. 

Using the information of TTT or EEE, we can estimate the size of $f_{\rm NL}^{sss}$ translated from the $sst$ correlator, equivalent to $f_{\rm NL}^{sst}$ in Eq.~\eqref{eq:fNLsst_theory}, according to $f_{\rm NL}^{sst} = \lambda_{sst} F_{sst, sss}^{XXX} / F_{sss,sss}^{XXX}$. Figure~\ref{fig:fNL} shows the results as a function of $\ell_{\rm max}$. For the TTT case, the $sst$ bispectrum is weakly correlated to the $sss$ one for $\ell_{\rm max} \gtrsim 100$ because, at such scales, the tensor-mode and scalar-mode transfer functions, ${\cal T}_{\ell(t)}^{T}(k)$ and ${\cal T}_{\ell(s)}^{T}(k)$, have very different shapes \cite{Pritchard:2004qp}. For $\ell_{\rm max} \lesssim 100$, however, the correlation recovers because of the shape similarity between the $sst$ and $sss$ bispectra due to the (Integrated) Sachs-Wolfe plateau \cite{Pritchard:2004qp}. For this reason, $|f_{\rm NL}^{sst}| $ falls below the theoretical expectation $\sim |\epsilon \lambda_{sst}|$ [see Eq.~\eqref{eq:fNLsst_theory}] for $\ell_{\rm max} \gtrsim 100$. For the EEE case, thanks to the weakness of the correlation between the $sst$ and $sss$ bispectra, $|f_{\rm NL}^{sst}|$ falls below $|\epsilon\lambda_{sst}| / 2$ for $\ell_{\rm max} \gtrsim 20$. As expected from Fig.~\ref{fig:blll}, if $\epsilon = 10^{-3}$, $f_{\rm NL}^{sst} / \lambda_{sst}$ with $N_{\rm tot} = 60$ almost overlaps that with $N_{\rm tot} = 0$.


\begin{figure}[t!]
  \begin{tabular}{cc}
    \begin{minipage}{0.5\hsize}
      \begin{center}
        $\epsilon = 10^{-3}$
    \includegraphics[width=1\textwidth]{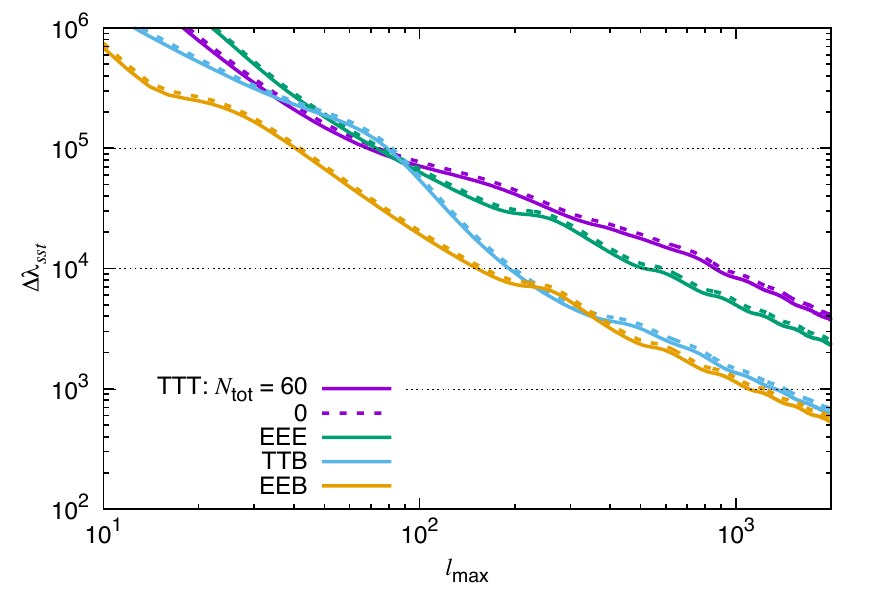}
  \end{center}
    \end{minipage}
    \begin{minipage}{0.5\hsize}
      \begin{center}
        $\epsilon = 10^{-2}$
    \includegraphics[width=1\textwidth]{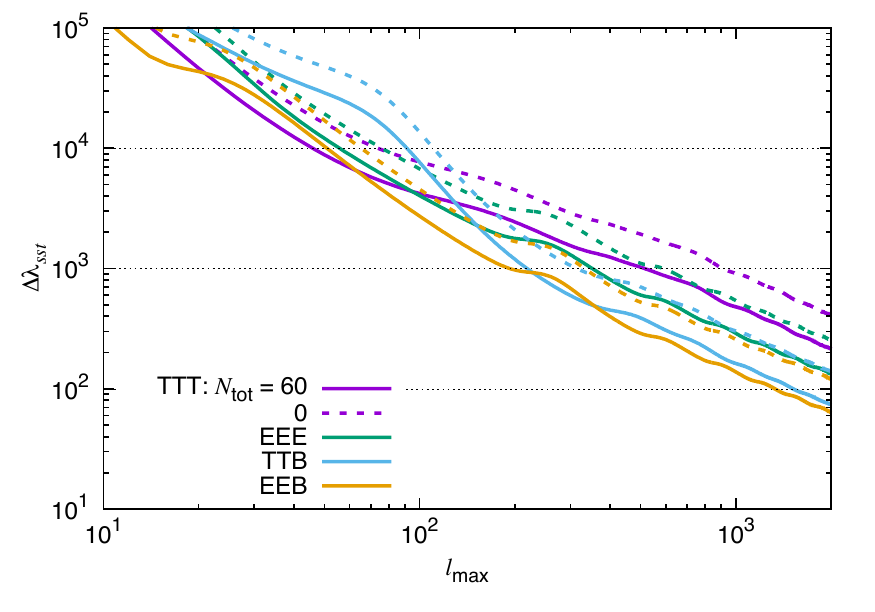}
  \end{center}
    \end{minipage}
    \end{tabular}
\\
\begin{tabular}{cc}
    \begin{minipage}{0.5\hsize}
  \begin{center}
    \includegraphics[width=1\textwidth]{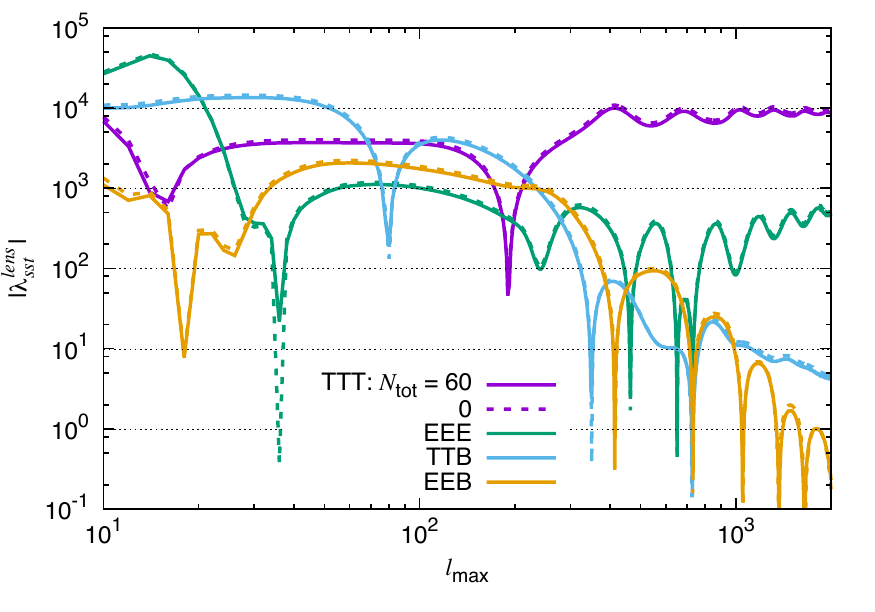}
  \end{center}
    \end{minipage}
    \begin{minipage}{0.5\hsize}
  \begin{center}
    \includegraphics[width=1\textwidth]{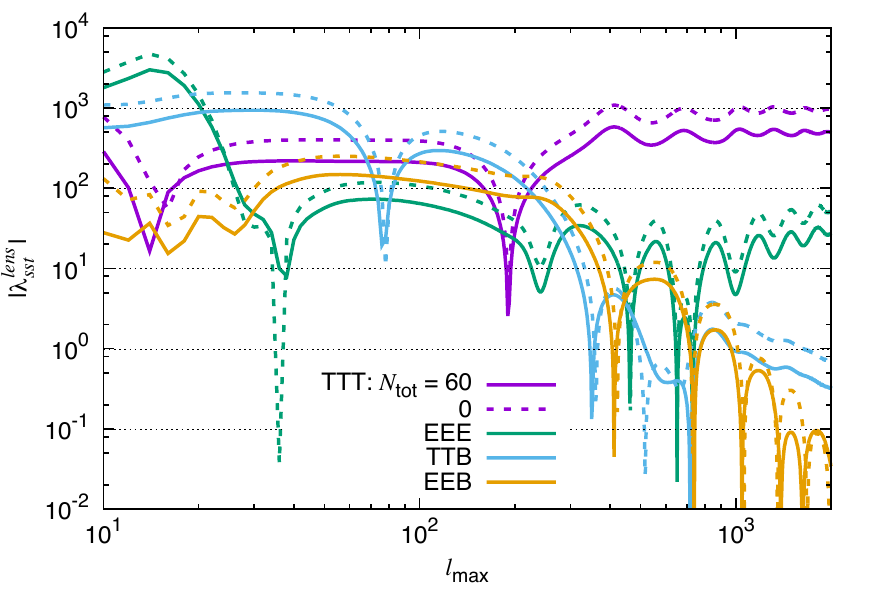}
  \end{center}
\end{minipage}
  \end{tabular}
  \caption{Expected $1\sigma$ errors on $\lambda_{sst}$ (top two panels) and the biases due to the secondary lensed bispectra on the estimation of $\lambda_{sst}$ (bottom two panels) with $\epsilon = 10^{-3}$ and $10^{-2}$, computed from the Fisher matrices of TTT, EEE, TTB and EEB. Solid and dotted lines correspond to the cases for $N_{\rm tot} = 60$ and $0$, respectively.}
\label{fig:error}
\end{figure}

The top panel of Fig.~\ref{fig:error} describes the expected 1$\sigma$ errors on $\lambda_{sst}$, computed according to $\Delta \lambda_{sst} = 1/\sqrt{F_{sst, sst}}$, in the TTT, EEE, TTB and EEB analyses. From this, we find a $\ell_{\rm max}$ scaling similar to the $sss$ case; namely $\Delta \lambda_{sst} \propto \ell_{\rm max}^{-1}$, in TTT and EEE \cite{Shiraishi:2010kd} and a bit more rapid scaling in TTB and EEB. In the case that the 2nd term of Eq.~\eqref{eq:tss1} contributes weakly to the Fisher matrix, Eq.~\eqref{eq:fish_XXX} indicates that $\Delta \lambda_{sst}$ from TTT or EEE simply scales like $\epsilon^{-1}$. Concerning TTB and EEB, in the Fisher matrix~\eqref{eq:fish_BXX}, the dominant contribution comes from the squeezed-limit signal: $\ell_1 \ll \ell_2 \sim\ell_3$. On such small $\ell_1$'s, $C_{\ell_1}^{BB}$ in the denominator is almost determined by the cosmic variance and thus $C_{\ell_1}^{BB} \propto r \propto \epsilon$ (since we are now considering not so small $r$). With a fact that $B_{\ell_1 \ell_2 \ell_3}^{BXX} \propto \epsilon$, we derive $\Delta \lambda_{sst} \propto \epsilon^{-1/2}$. These relations are confirmed from the lines except for $(\epsilon, N_{\rm tot}) = (10^{-2}, 60)$. Because of this, $\Delta \lambda_{sst}$ at $\ell_{\rm max} = 2000$ depends strongly on $\epsilon$, such as $10^3 - 10^4$ ($10^2 - 10^3$) for $\epsilon = 10^{-3}$ ($10^{-2}$). Note that our $\Delta \lambda_{sst}$ for $N_{\rm tot} = 0$ obtained from TTT (TTB) is in agreement with the corresponding results reported in Ref.~\cite{Shiraishi:2010kd} (Ref.~\cite{Meerburg:2016ecv}).

It is important to investigate how much the late-time secondary contributions contaminate the primordial signal discussed above. We now consider the TTT, EEE, TTB and EEB bispectra induced by the primordial temperature and E-mode signal via late-time gravitational lensing, which are the dominant components of the secondary contributions.
 Because of the smallness of the primordial BB spectrum and the absence of the parity-violating primordial correlators such as TB or EB, the leading-order expressions are given by the products of the lensed TT/TE/EE spectrum and the temperature/E-mode polarization-lensing potential cross power spectrum \cite{Hu:2000ee,Lewis:2011fk}. In the bottom panel of Fig.~\ref{fig:error} we describe the biases due to such lensed contributions on the estimation of $\lambda_{sst}$, computed according to $\lambda_{sst}^{lens} = F_{sst, lens} / F_{sst,sst}$. From the TTT result, we find that $|\lambda_{sst}^{lens}|$ surpasses $\Delta \lambda_{sst}$ for $\ell_{\rm max} \gtrsim 1000$. This suggests a caveat against the $\lambda_{sst}$ estimation without the subtraction of the secondary lensed contributions in the case that one uses the information of TTT beyond $\ell \simeq 1000$. In contrast, there will be no such a concern in the data analysis with EEE, TTB or EEB (at least up to $\ell = 2000$) because of $|\lambda_{sst}^{lens}| \ll \Delta \lambda_{sst}$ for all $\ell_{\rm max}$'s.


\newcommand{\kk}{\bm{k}}
\newcommand{\rr}{\bm{r}}
\newcommand{\mpl}{M_{\rm pl}}
\newcommand{\pending}{{\color{red}[PENDING]}}

\section{Conclusion and Discussion}
\label{sec:cd}

Dynamics of the standard, single-field slow-roll inflation is now 
 well understood including the properties of the linear perturbation,
and the theoretical predictions, i.e. the 2-point function or the power spectrum, 
the spectral index, and the tensor-to-scalar ratio have been tested against
the observed CMB data, which resulted in a rather tight constraint on the
single-field slow-roll models.

In contrast, the non-linear dynamics is still far from complete
understanding. It is expected that it is the key to learn more about 
the theory of gravity in the early Universe, in particular 
through the 3-point functions.  In this connection, the latest reported 
Planck results on non-Gaussianities lead us to speculate that 
actually non-Gaussianities, in particular the local-type ones,
might be scale dependent. If this was the case, all models of single-field 
slow-roll inflation would be excluded because the main contribution to 
the bispectrum in those models comes from the cubic scalar interaction
which must be first of all very small, and which wouldn't produce
such scale-dependence. Of course, one may achieve such a feature
by resorting to multifield models, with some tuning of the parameters.

In this work, we proposed an alternative scenario. We proposed a model
in which the bispectrum is mainly sourced by the almost scale-invariant primordial
scalar-scalar-tensor interactions during inflation. We achieved a sizeable 
interaction, actually larger than the one from the cubic scalar interaction,
by considering a model in which there exists a preferred spatial frame,
which breaks the local SO(3) symmetry (part of the local Lorentz symmetry).
This is realized by introducing three scalar fields which turn to be
non-dynamical and give rise to a preferred spatial frame.
It leads to an effective mass for the tensor modes and to an 
enhancement of the interaction between scalar and tensor modes. 
We showed that it generates scale-dependent non-Gaussianities similar to 
what WMAP and Plank probed at low $\ell$.
The scale dependence in the CMB bispectrum is naturally obtained 
because the tensor model transfer function is strongly scale-dependent 
due to the decay of the tensor mode after it re-enters the horizon.

We parametrized the amplitude of the interaction term and, hence, of 
the non-Gaussianity with the parameter $\lambda_{sst}$, Eq.\eqref{lamda}. 
We found that $\lambda_{sst} \epsilon >1$ is required in order to
 explain the current amplitude at low $\ell$ multipoles. 
We performed a thorough study of the effects of this 3-point function on 
the temperature and polarization bispectra. From the Fisher matrix analysis 
based on an accurate all-sky CMB formalism, we found that,
 a minimum detectable $\lambda_{sst}$ from TTT (EEE) up 
to $\ell = 2000$ is $\simeq 4 / \epsilon$ ($2 / \epsilon$), and it could 
be improved by more than one order of magnitude if using the information 
of TTB or EEB. It is therefore interesting to analyze both the current 
temperature/E-mode maps in WMAP and Planck and the B-mode one in 
the futuristic CMB experiments with the shape and scale dependence
 of non-Gaussianity that we have proposed.

\section*{Acknowledgement}
This work was supported in part by
the MEXT KAKENHI Nos.~15H05888 and 15K21733.
T. Hiramatsu is supported by JSPS Grant-in-Aid for
Scientific Research on Innovative Areas No.~16H01098, and No.~15H05888. 
C. Lin is supported by JSPS postdoc fellowship for overseas researchers,
and by JSPS Grant-in-Aid for Scientific Research No~.15F15321.
M. Shiraishi is supported in part by a Grant-in-Aid for JSPS Research under 
Grant No.~27-10917, and in part by the World Premier International Research 
Center Initiative (WPI Initiative), MEXT, Japan. Numerical computations by 
M. Shiraishi were in part carried out on Cray XC30 at Center for 
Computational Astrophysics, National Astronomical Observatory of Japan. 
Y. Wang is supported by the CRF Grants of the Government of the Hong Kong 
SAR under HKUST4/CRF/13G and ECS 26300316.
Y. Wang would like to thank the Yukawa Institute for Theoretical Physics 
for hospitality where part of this work was done. 


\appendix

\section{On the gauges}
\label{appsec:details}
In this appendix, we give some explicit formulas which might be useful
to reproduce the main part of the paper.
At the same time, we show that the scalar-scalar-tensor interaction 
we consider in our model is not a gauge artifact. 
 Form the definition of
$Z^{ij}$, equation \eqref{eq:Z}, we have that up to the second order in perturbations
\begin{align}
Z^{ij}=h^{ij}+h^{(ik}\partial_k\pi^{j)}+\partial_k\pi^i\partial_k\pi^j-\left(N^i+\dot\pi^i\right)\left(N^j+\dot\pi^j\right)\,.
\end{align}
It is instructive to see that at the leading order we have
\begin{align}
3\bar{\delta} Z^{ij}\simeq\gamma_{ij}+\partial_i\partial_j(E-\pi)+O(2)\,,
\end{align}
where we decomposed $\pi^i=\pi^i_T+\partial^i\pi$ with
$\partial_i\pi^i_T=0$, thus leading to the second order action
\eqref{s2}. Most importantly, we see that $E$ and $\pi$ appear in a
gauge invariant combination \cite{Gumrukcuoglu:2011zh}, that is
$E_\pi=E-\pi$. The first order constraint yields $E_\pi=0$. Let us fix
the spatial gauge degrees of freedom to $E=\pi=0$ for simplicity. Thus,
up to second order we obtain
\begin{align}
\begin{split}
3\bar\delta Z^{ij}&=\gamma_{ij}-\frac{3}{2}\gamma_{ik}\gamma_{kj}+\frac{1}{6}\delta_{ij}\gamma_{kl}\gamma_{kl}
+a^{2}\partial_i\beta\partial_j\beta-\frac{a^{2}}{3}\delta_{ij}\partial_k\beta\partial_k\beta+O(3)\,.
\end{split}
\end{align}
From this expression, it is easy to reproduce the third order action
\eqref{eq:s3}.
Let us recall that in the spatially isotropic comoving slicing gauge
one has \cite{Kodama:1985bj, Maldacena:2002vr, Koyama:2010xj}
\begin{align}
\beta_{com}=-\frac{\cR}{H}+a^2\epsilon\partial^{-2}\dot\cR
\end{align}
and in the spatially isotropic flat slicing gauge,
\begin{align}
\beta_{flat}=-\frac{H}{\dot\phi}a^2\epsilon\left(\partial^{-2}\dot\delta\phi+(\epsilon+\delta)H\partial^{-2}\delta\phi\right)\,,
\end{align}
where $\delta=\ddot\phi/(H\dot\phi)$.
Regarding the interaction term, once the spatial gauge is fixed, it is given by
\begin{align}\label{eq:interaction}
{\cal L}_{sst}\supset\bar\delta Z^{ij}\partial^\mu\varphi^i\partial_\mu\phi\,\partial^\nu\varphi^j\partial_\nu\phi=\frac{\dot\phi}{3}\gamma_{ij}\,\partial_i\left(\beta+\delta\phi/\dot\phi\right)\partial_j\left(\beta+\delta\phi/\dot\phi\right)\,.
\end{align}
In particular, notice that it appears in a temporal gauge invariant 
combination \cite{Kodama:1985bj,Mukhanov:1990me}. This shows the
existence of the scalar-scalar-tensor interaction irrespective 
of the gauge.




\end{document}